\documentclass[reprint,superscriptaddress,aps,twocolumn,showpacs]{revtex4-1}
\usepackage{amssymb}
\usepackage{amsmath}
\usepackage{graphicx}
\usepackage{multirow}
\usepackage{hyperref}

\allowdisplaybreaks

\begin{document}
\title{Study of charmoniumlike and fully-charm tetraquark spectroscopy}

\author{Zheng Zhao}
\email[]{zhaozheng1022@hotmail.com}
\author{Kai Xu}
\author{Attaphon Kaewsnod}
\affiliation{School of Physics and Center of Excellence in High Energy Physics and Astrophysics, Suranaree University of Technology, Nakhon Ratchasima 30000, Thailand}
\author{Xuyang Liu}
\affiliation{School of Physics and Center of Excellence in High Energy Physics and Astrophysics, Suranaree University of Technology, Nakhon Ratchasima 30000, Thailand}
\affiliation{School of Physics, Liaoning University, Shenyang 110036, China}
\author{Ayut Limphirat}
\author{Yupeng Yan}
\email[]{yupeng@sut.ac.th}
\affiliation{School of Physics and Center of Excellence in High Energy Physics and Astrophysics, Suranaree University of Technology, Nakhon Ratchasima 30000, Thailand}

\date{\today}

\begin{abstract}
\indent The masses of tetraquark states of all $qc\bar q \bar c$ and $cc\bar c \bar c$ quark configurations are evaluated in a constituent quark model, where the Cornell-like potential and one-gluon exchange spin-spin coupling are employed.
All model parameters are predetermined by comparing the theoretical and experimental masses of light, charmed and bottom mesons.
The theoretical predictions of the charmoniumlike tetraquarks are compared with the observed $XYZ$ states, and one tentative assignment is suggested. The work suggests that the $X(6900)$ observed by LHCb is likely to be the first radial excited fully-charm tetraquark state with $J^{PC} = 1^{+-}$ in the $\bar 3_c \otimes 3_c$ configuration, and the ground and second radial excited states of fully-charm tetraquark are around $6494$ and $7253$ MeV respectively.
\end{abstract}

\maketitle

\section{Introduction}\label{sec:Int}

\indent The possible existence of multiquarks like tetraquarks ($q^2\bar q^2$), pentaquarks ( $q^4\bar q$ ), dibaryon ($q^6$) and baryonium ($q^3\bar q^3$) and that baryons could be the combination of $q^3$ and $q^4\bar q$ while mesons could be made out of $q\bar q$ and $q^2\bar q^2$ were suggested at the birth of the quark model \cite{GellMann:1964nj, Zweig:1964jf}.
In the past two decades, over 20 charmoniumlike and bottomoniumlike XYZ states have been observed\cite{Brambilla:2019esw}, which do not fit into the naive quark model spectrum easily. The observations of the XYZ states have inspired extensive interests of theorists in revealing their underlying structures. Listed in Table~\ref{tab:particle} are the masses, widths and $J^{PC}$ of these XYZ states in the $c\bar c$ region from the listed sources, or are taken from Particle Data Group(PDG) \cite{PDG} when available. Instead of using the latest PDG naming scheme, we shall use $X$ for all neutral states, $Z_c$ for all charged states, and $Y$ for those $J^{PC}=1^{--}$ states throughout this paper.

\indent A number of charged charmoniumlike particles like $Z^+_c(3900)$, $Z^+_c(4020)$, $Z^+(4050)$, $Z^+(4055)$, $Z^+_c(4100)$, $Z^+_c(4200)$, $Z^+_c(4250)$ and $Z^+_c(4430)$ have been successively observed by experimental collaborations \cite{Brambilla:2019esw}. Obviously, those charged charmoniumlike states go beyond the conventional $c\bar c$-meson picture but are likely tetraquark systems $c\bar cu\bar d$ due to carrying one charge, which provides a good place for testing various phenomenological research methods of hadron physics.
A systematic understanding of the internal structure of the charged charmoniumlike $Z_c$ states would provide new insights into the dynamics of multiquark systems as well as information for the future experimental search for the missing higher excitations in the $Z^+_c$ family.
\begin{table*}[t]
\caption{\label{tab:particle} Masses, widths and $J^{PC}$ of X, Y and Z states in the $c\bar c$ region from the cited sources or \cite{PDG} when available.}
\begin{ruledtabular}
\begin{tabular}{lcccccc}
States &Name in PDG&M(MeV)&$\Gamma$&  $J^{PC}$ &  Process & Experiment
\\
\hline
$X(3860)$  &$\chi_{c0}(3860)$ &  $3862^{+26+40}_{-32-13}$  &  $201^{+154+88}_{-67-82}$  &  $  0^{++}$  &  $e^+e^-\to J/\psi (D\bar D)$ & Belle\cite{Chilikin:2017evr}
\\
$X(3915)$  &$X(3915)$&  $3918.4\pm 1.9$  &  $20\pm5$  &  $0/2^{++}$  &  $B\to K(J/\psi \omega)$ & Belle\cite{Abe:2004zs}
\\
$X(3940)$  &$X(3940)$&  $3942^{+7}_{-6}\pm 6$  &  $37^{+26}_{-15}\pm18$  &  $?^{??}$  &  $e^+e^-\to J/\psi (D\bar D^*)$  &  Belle\cite{Abe:2007sya}
\\
$X(4160)$  &$X(4160)$&  $4156^{+25}_{-20}\pm 15$  &  $139^{+111}_{-61}\pm21$  &  $?^{??}$  &  $e^+e^-\to J/\psi (D^*\bar D^*)$  &  Belle\cite{Abe:2007sya}
\\
$X(4350)$  &$X(4350)$&  $4350.6^{+4.6}_{-5.1}\pm 0.7$  &  $13^{+18}_{-9}\pm4$  &  $?^{?+}$  &  $\gamma \gamma \to \phi J/\psi $  &  Belle\cite{Shen:2009vs}
\\
$X(4500)$ & $\chi_{c0}(4500)$ &  $4506\pm{11}^{+12}_{-15}$  &  $92\pm21^{+21}_{-20}$  &  $0^{++}$  &  $B^+ \to (J/\psi \phi) K^+ $  &  LHCb\cite{Aaij:2016iza}
\\
$X(4700)$ &  $\chi_{c0}(4700)$  &  $4704\pm{10}^{+14}_{-24}$  &  $120\pm31^{+42}_{-33}$  &  $0^{++}$  &  $B^+ \to (J/\psi \phi) K^+ $  &  LHCb\cite{Aaij:2016iza}
\\
\hline
$Y(4230)$  &$\psi(4230)$&  $4218^{+5}_{-4}$&  $59^{+12}_{-10}$  &  $1^{--}$  &  $e^+e^- \to \omega\chi_{c0}$  &  BESIII\cite{Ablikim:2014qwy}
\\
$Y(4260)$  &$\psi(4260)$&  $4230\pm8$&  $55\pm19$  &  $1^{--}$  &  $e^+e^- \to \pi^+\pi^-J/\psi$  &  BaBar\cite{Aubert:2005rm}
\\
$Y(4360)$ & $\psi(4360)$ &  $4368\pm{13}$&  $96\pm7$  &  $1^{--}$  &  $e^+e^- \to\pi^+\pi^-\psi(2S)$  & BESIII\cite{Ablikim:2017oaf}
\\
$Y(4390)$  &$\psi(4390)$ &  $4391.5^{+6.3}_{-6.8}\pm{1.0}$&  $139.5^{+16.2}_{-20.6}\pm{0.6}$  &  $1^{--}$  &  $e^+e^- \to  h_c\pi^+\pi^- $  &  BESIII\cite{BESIII:2016adj}
\\
$Y(4660)$  & $\psi(4660)$ &  $4643\pm{9}$&  $72\pm11$  &  $1^{--}$  &  $e^+e^- \to\pi^+\pi^-\psi(2S) $  & Belle\cite{Wang:2014hta}
\\
\hline
$Z_c(3900)$ &$ Z_c(3900)$ &  $3888.4\pm{2.5}$  &  $28.3\pm2.5$  &  $1^{+-}$  &  $e^+e^- \to (D\bar D^*)^+\pi^- $ & BESIII\cite{Ablikim:2013xfr}
 \\
$Z_c(4020)$  &$X(4020)^{\pm}$&  $4024.1\pm{1.9}$  &  $13\pm5$  &  $?^{?-}$  &  $e^+e^- \to\pi^-(\pi^+h_c) $ & BESIII\cite{Ablikim:2013wzq}
\\
$Z_c(4050)$  &$X(4050)^{\pm}$&  $4051^{+24}_{-40}$  &  $82^{+50}_{-28}$  &  $?^{?+}$  &  $\bar B^0 \to K^-(\pi^+\chi_{c1}) $  &  Belle\cite{Mizuk:2008me}
\\
$Z_c(4055)$  &$X(4055)^{\pm}$&  $4054\pm{3.2}$&  $  45\pm13$  &  $?^{?-}$  &  $e^+e^- \to\pi^-(\pi^+\psi(2S)) $  & Belle\cite{Wang:2014hta}
\\
$Z_c(4100)$  &$X(4100)^{\pm}$&  $4096\pm{28}$  &  $152^{+80}_{-70}$  &  $0^{++}/1^{-+}$  &  $B^0\to K^+(\pi^-\eta_c) $  &  LHCb\cite{Aaij:2018bla}
 \\
$Z_c(4200)$ & $Z_c(4200)$  &  $4196^{+35}_{-32}$  &  $370^{+100}_{-150}$  &  $1^{+-}$  &  $\bar B^0\to K^-(\pi^+J/\psi) $  &  Belle\cite{Chilikin:2014bkk}
\\
{$Z_c(4250)$}  &$X(4250)^{\pm}$&  $4248^{+190}_{-50}$  &  $177^{+320}_{-70}$  &  $?^{?+}$  &  $\bar B^0\to K^-(\pi^+\chi_{c1})$  &  Belle\cite{Mizuk:2008me}
\\
$Z_c(4430)$ & $Z_c(4430)$ &  $4478^{+15}_{-18}$  &  $181\pm31$  &  $1^{+-}$  &  $\bar B^0\to K^-(\pi^+J/\psi)$ & Belle\cite{Chilikin:2014bkk}
\end{tabular}
\end{ruledtabular}
\end{table*}

\indent In the past decades, a number of models and improved pictures have been proposed to study the tetraquark states. We briefly discuss some representatives here and later in Sec.~\ref{sec:DIS}.

\indent The X(3872) was interpreted as a diquark-antidiquark charmoniumlike tetraquark state, and a spectrum was derived by taking the X(3872) as input\cite{Maiani:2004vq}. The model was named "Type-I" diquark-antidiquark model. In this model, the diquark constituent quark masses were fixed by inputting the X(3872) state, and other parameters were fixed by the conventional meson spectrum and conventional baryon mass spectrum, where each diquark or quark-antiquark pair has its own parameters.  Later, the $Z_c(3900)$ state was interpreted as a diquark-antidiquark charmoniumlike tetraquark state and its decay modes were investigated\cite{Faccini:2013lda}. The "Type-I" diquark-antidiquark model was further developed, and a "Type-II" diquark-antidiquark model was proposed, which has more complicated spin-spin interactions \cite{Maiani:2014aja}.

\indent The diquark-antidiquark picture was applied to study the $Z_c(3900)$/$Z_c(3885)$ and $Z_c(4020)$/$Z_c(4025)$ states\cite{Patel:2014vua}, where the parameters were fitted by inputting the mass of $X(3823)$, $Z_c(3900)$ and $Z_c(3885)$ states. In Ref.\cite{Patel:2014vua}, the $Z_c(3900)$ and $Z_c(4025)$ were assigned as $Q\bar q-\bar Qq$ molecularlike four quark states, and the $Z_c(3885)$ as a diquark-antidiquark tetraquark state.

\indent Within the framework of the color-magnetic interaction, the mass spectra of the hidden-charm and hidden-bottom tetraquark states were studied together \cite{Zhao:2014qva}, where the constituent quark mass was fixed by charmonium meson spectrum, but the other parameters were fixed by inputting the masses of newly observed $Z_c(3900)$, $Z_c(4025)$ and $Z_c(4200)$ states.

\indent In this work we will not pay much attention to repeating the mass of potential tetraquark states by applying a large number of parameters but apply a simple model, where all the model parameters are predetermined by the conventional mesons, to predict the mass of all possible tetraquark configurations.
The paper is organized as follows. In Sec.~\ref{sec:CWF}, we work out all the possible configurations of color, flavor, spin, and spatial degrees of freedom of tetraquark states.
The constituent quark model applied in our previous work \cite{Xu:2019fjt} is briefly reviewed in Sec.~\ref{sec:GTM}, and all model parameters are determined by comparing the theoretical and experimental masses of light, charmed and bottom mesons.
In Sec.~\ref{sec:PTM}, tetraquark mass spectra are evaluated in the constituent quark model with all parameters predetermined.
In Sec.~\ref{sec:DIS}, the theoretical results are compared with experimental data and one tentative assignment for charmoniumlike tetraquarks is suggested.
A summary is given in Sec.~\ref{sec:SUM}. The details of the construction of tetraquark wave functions are shown in the Appendix.

\section{\label{sec:CWF}Construction of wave function}
\subsection{Quark configurations of tetraquark}

\indent The construction of tetraquark states follows the rule that a tetraquark state must be a color singlet, that is, the tetraquark color wave function must be a $[222]_1$ singlet of the $SU_c(3)$ group. The permutation symmetry of the two quarks cluster ($qq$) of tetraquark states is characterized by the Young tabloids $[2]_6$, $[11]_{\bar 3}$ of the the $SU_c(3)$ group while the color part of the two antiquarks cluster ($\bar q\bar q$) is a $[211]_3$ triplet and $[22]_{\bar 6}$ antisextet, thus a $[222]_1$ color singlet of tetraquark states demands the following configurations:
\begin{flalign}\label{eq:colorcon}
&[2]_6(q_1q_2)\otimes[22]_{\bar 6}(\bar q_3\bar q_4),\nonumber \\
&[11]_{\bar 3}(q_1q_2)\otimes[211]_3(\bar q_3\bar q_4),
\end{flalign}
with the corresponding dimensions being: $\bar 3_c \otimes 3_c$ and $6_c \otimes \bar 6_c$.

\indent For tetraquark states of four light quarks or four charm quarks, a $qq\bar q\bar q$ state must be a color singlet and antisymmetric under any permutation between identical quarks, which  implies that the spatial-spin-flavor part of color part $[2]_6$ and $[11]_{\bar 3}$ should be [11] and [2] states by conjugation, respectively.
Listed in Table~\ref{tab:ssf} are all the possible configurations of spatial-spin-flavor part of the $qq$ cluster.
\begin{table}[tb]
\caption{All possible color-spatial-spin-flavor configurations of $qq$}
\label{tab:ssf}
\begin{ruledtabular}
\begin{tabular}{lcc}
\phantom{} & $\psi^c_{[2]}\psi^{osf}_{[11]}$ & \phantom{}
\\
$\psi^c_{[2]}\psi^o_{[2]}\psi^s_{[11]}\psi^f_{[2]}$ & \phantom{} & $\psi^c_{[2]}\psi^o_{[2]}\psi^s_{[2]}\psi^f_{[11]}$
\\
$\psi^c_{[2]}\psi^o_{[11]}\psi^s_{[11]}\psi^f_{[11]}$ & \phantom{} & $\psi^c_{[2]}\psi^o_{[11]}\psi^s_{[2]}\psi^f_{[2]}$
\\
\hline
\phantom{} & $\psi^c_{[11]}\psi^{osf}_{[2]}$ & \phantom{}
\\
$\psi^c_{[11]}\psi^o_{[11]}\psi^s_{[2]}\psi^f_{[11]}$ & \phantom{} & $\psi^c_{[11]}\psi^o_{[11]}\psi^s_{[11]}\psi^f_{[2]}$
\\
$\psi^c_{[11]}\psi^o_{[2]}\psi^s_{[11]}\psi^f_{[11]}$ & \phantom{} & $\psi^c_{[11]}\psi^o_{[2]}\psi^s_{[2]}\psi^f_{[2]}$
\end{tabular}
\end{ruledtabular}
\end{table}

\indent In this work we consider only charmoniumlike ($qc\bar q \bar c$) and fully-charm tetraquark ($cc\bar c\bar c$) states.
The color wave function of each tetraquark color configuration may be written in the general form
\begin{flalign}\label{eq:color}
&\psi^{qc\bar q\bar c}_{[2]_{6}^c{{[22]}^c_{\bar 6}}}=\frac 1{\sqrt 6}\sum_{i=1}^{6}\psi_{[2]_6^ci}^{qc}\psi_{[22]_{\bar 6}^ci}^{\bar q \bar c},\nonumber \\
&\psi^{qc\bar q\bar c}_{[11]_{\bar 3}^c{{[211]}^c_{3}}}=\frac 1{\sqrt 3}\sum_{i=1}^{3}\psi_{[11]_{\bar 3}^ci}^{qc}\psi_{{[211]}^c_{3}i}^{\bar q \bar c}
\end{flalign}
which are for the color sextet-antisextet ($6_c \otimes \bar 6_c$) and triplet-antitriplet ($\bar 3_c \otimes 3_c$) configurations, respectively.
The explicit color wave functions are listed in the Appendix~\ref{sec:AP1}. The possible spin combinations are 
$\left[\psi_{[s=1]}^{qc}\otimes\psi_{[s=1]}^{\bar q \bar c}\right]_{S=0,1,2}$, $\psi_{[s=1]}^{qc}\otimes\psi_{[s=0]}^{\bar q \bar c}$,
$\psi_{[s=0]}^{qc}\otimes\psi_{[s=1]}^{\bar q \bar c}$, and $\psi_{[s=0]}^{qc}\otimes\psi_{[s=0]}^{\bar q \bar c}$. 
\begin{flalign}\label{eq:spin}
& \left[\psi_{[s=1]}^{qc}\otimes\psi_{[s=1]}^{\bar q \bar c}\right]_{S=0,1,2}, \nonumber \\
& \psi_{[s=1]}^{qc}\otimes\psi_{[s=0]}^{\bar q \bar c},\;\; \psi_{[s=0]}^{qc}\otimes\psi_{[s=1]}^{\bar q \bar c}, \nonumber \\
& \psi_{[s=0]}^{qc}\otimes\psi_{[s=0]}^{\bar q \bar c}. 
\end{flalign}

For $cc\bar c\bar c$ tetraquarks, the spin wave functions for $cc$ and $\bar c\bar c$ must be symmetric and antisymmetric are for $[11]_{\bar 3}$ and $[2]_6$ color configurations, respectively.  The explicit spin wave functions for $cc\bar c\bar c$ are listed in the Appendix~\ref{sec:AP1}

\subsection{Spatial wave function}

\indent We construct the complete bases by using the harmonic oscillator wave function. The corresponding symmetries of the spatial wave functions are enforced to guarantee the antisymmetric property of identical particles.

\indent The relative Jacobi coordinates and the corresponding momenta are defined as
\begin{flalign}\label{eqn::ham}
&\vec x_1=\frac{1}{\sqrt 2}(\vec r_1-\vec r_3), \nonumber \\
&\vec x_2=\frac{1}{\sqrt 2}(\vec r_2-\vec r_4), \nonumber \\
&\vec x_3=\frac{m_1\vec r_1+m_3\vec r_3}{m_1+m_3}-\frac{m_2\vec r_2+m_4\vec r_4}{m_2+m_4}, \nonumber \\
&\vec x_0=\frac{m_1\vec r_1+m_2\vec r_2+m_3\vec r_3+m_4\vec r_4}{m_1+m_2+m_3+m_4}, \nonumber \\
&\vec{p}_{i}= u_i \frac{d\vec x_i}{dt},
\end{flalign}
where $u_i$ are the reduced quark masses defined as
\begin{flalign}\label{eqn::rqm}
&u_1=\frac{2m_1m_3}{m_1+m_3}, \nonumber \\
&u_2=\frac{2m_2m_4}{m_2+m_4}, \nonumber \\
&u_3=\frac{(m_1+m_3)(m_2+m_4)}{m_1+m_2+m_3+m_4},
\end{flalign}
where $\vec{r}_{j}$ and $m_j$ are the coordinate and mass of the jth quark. We assign $x_1$, $x_2$, $x_3$ to be $\sigma_1$, $\sigma_2$ and $\lambda$ Jacobi coordinates, respectively.

\indent The total spatial wave function of tetraquark, coupling among the $\sigma_1$, $\sigma_2$ and $\lambda$ harmonic oscillator wave functions, may take the general form,
\begin{eqnarray}\label{eqn::spatial}
\psi_{NLM} &=&
\sum_{\{n_i,l_i\}}
 A(n_{\sigma_1},n_{\sigma_2},n_{\lambda},l_{\sigma_1},l_{\sigma_2},l_{\lambda}) \nonumber \\
&& \times \psi_{n_{\sigma_1}l_{\sigma_1}}(\vec\sigma_1\,) \otimes\psi_{n_{\sigma_2}l_{\sigma_2}}(\vec\sigma_2\,)\otimes\psi_{n_{\lambda}l_{\lambda}}(\vec\lambda\,)
\end{eqnarray}
where $\psi_{n_{i}l_{i}m_{i}}$ are just harmonic oscillator wave functions and the sum $\{n_i,l_i\}$ is over $n_{\sigma_1},n_{\sigma_2},n_{\lambda}, l_{\sigma_1},l_{\sigma_2},l_{\lambda}$. $N$, $L$, and $M$ are respectively the total principle quantum number, total angular momentum, and magnetic quantum number of tetraquark. One has $N= (2n_{\sigma_1}+ l_{\sigma_1})+(2n_{\sigma_2}+ l_{\sigma_2})+(2n_{\lambda}+l_{\lambda})$. 

We will employ the spatial wave functions $\psi_{NLM}$, grouped according to the permutation symmetry, as complete bases to study tetraquark states with other interactions. In our calculations, the bases size is N=18, and the length parameter of harmonic oscillator wave functions is adjusted to 450 MeV to get the best eigenvalue of Eq. (\ref{eqn::se}). The complete bases of the tetraquarks are listed in the Appendix~\ref{sec:AP1}, Table \ref{three1}, up to $N=10$, where $l_{\sigma_1}$, $l_{\sigma_2}$, and $l_{\lambda}$ are limited to $0$ only.

\section{\label{sec:GTM}THEORETICAL MODEL}

\begin{table}[b]
\caption{Spin-averaged masses $M_k^{ave}$ for various kinds of mesons. $M_{PS}$ and $M_V$ taken from PDG~\cite{PDG}.}
\label{tab:Maverage}
\begin{ruledtabular}
\begin{tabular}{lcccccccccc}
Meson & $ c\bar c$ & $b\bar b $ & $B_s$ & $B$ & $D_s$ & $D$ & $s\bar s$ & $q\bar q$
\\
$[MeV]$&3068&9444&5404&5314&2076&1972&952&675
\\
\end{tabular}
\end{ruledtabular}
\end{table}

\indent We study the meson and tetraquark systems in the nonrelativistic Hamiltonian,
\begin{flalign}\label{eqn::ham}
&H =H_0+ H_{hyp}^{OGE}, \nonumber \\
&H_{0} =\sum_{k=1}^{N} (\frac12M^{ave}_{k}+\frac{p_k^2}{2m_{k}})+\sum_{i<j}^{N}(-\frac{3}{16}\lambda^{C}_{i}\cdot\lambda^{C}_{j})(A_{ij} r_{ij}-\frac{B_{ij}}{r_{ij}}),  \nonumber \\
&H_{hyp} =\sum_{i<j}C_{ij}{\lambda^{C}_{i}\cdot\lambda^{C}_{j}}\,\vec\sigma_{i}\cdot\vec\sigma_{j},
\end{flalign}
by solving the $\rm Schr\ddot{o}dinger$ equation
\begin{flalign}\label{eqn::se}
H|\psi \rangle =E|\psi \rangle, 
\end{flalign}
where the tetraquark wave function $|\psi \rangle$ is expanded in the complete bases defined in Eq. (\ref{eqn::se}). Here $m_k$ are the constituent quark masses. $M^{ave}_k$ denotes the spin-averaged mass as $\frac{1}{4}M_{PS}+\frac{3}{4}M_V$ (except for $s\bar s$ and $q\bar q$), with $M_{PS}$ and $M_V$ being the mass of ground state pseudoscalar and vector mesons from experimental data. The spin-averaged masses $M_k^{ave}$ for each kind of mesons are listed in Table~\ref{tab:Maverage}. The spin-averaged masses $M_k^{ave}$ for $s\bar s$ and $q\bar q$ are fitted by comparing theoretical results and experimental data to avoid the would-be Goldstone bosons of the chiral symmetry breaking\cite{Brambilla:2019esw}.

The Cornell potential 
\begin{flalign}\label{eqn::cp}
V(r)=Ar-\frac Br 
\end{flalign}
has been widely employed in quark model studies and also in lattice QCD simulations. Usually, the coefficient of string tension $A$ and Coulomb coefficient $B$ are fitted to experimental data. In quark model studies, model parameters are different for different hadrons , which also happens in lattice QCD studies \cite{Ikeda:2011bs, Kawanai:2011xb}.

The average velocity of large quark mass is predicted by the Cornell potential to saturate at the value $\langle v^2\rangle=B^2$, where B is the Coulomb coefficient in Cornell potential. According to the approximate equality of bottomonia and charmonia level splittings, $\langle v^2\rangle_{\Upsilon}/\langle v^2\rangle_{J/\psi}\approx m_c/m_b$ would be expected \cite{Bali:2000gf}. Thus, the Coulomb coefficients are mass dependent for large quark mass take the form $B_{b\bar b}=B_{c\bar c}\sqrt{{m_c}/{m_b}}$.

In Ref \cite{Kawanai:2011xb}, interquark potentials ($V_{q\bar q}$) at finite quark mass have been studied for heavy quarkonia in lattice QCD. The Cornell potential is employed for fitting the data. The fit results of Coulomb coefficient are massdependent with the finite quark mass ($m_{i}$) from 1.0 to 3.6 GeV, which take the form $B=B_0\sqrt{1/{m_{i}}}$. 

Later, another interquark potential study for $q\bar q$ systems has been done in a quenched lattice QCD \cite{Ikeda:2011bs}. One of the fitting functions is of Cornell-type, $V(r) = Ar-B/r+C$. The summarized fitting results show that Coulomb coefficients are massdependent with constant quark masses $m_q$ which are from $0.52$ to $1.275$ GeV and determined by half of vector meson masses $M_V$, i.e., $m_q = M_V /2$. The massdependent Coulomb coefficients take the form $B=B_0\sqrt{1/{m_{q}}}$. Meanwhile, the coefficient of string tension are linear massdependent with the form, $A=a+b\ {m_q}$.

\begin{table}[tb]
\caption{Color matrix elements of tetraquarks}
\label{tab:cme}
\begin{ruledtabular}
\begin{tabular}{lcc}
$\hat{O}$ & $<\psi^c_{6-\bar 6}|\hat{O}|\psi^c_{6-\bar 6}>$ & $<\psi^c_{\bar 3-3}|\hat{O}|\psi^c_{\bar 3-3}>$\\
\hline
$ \vec \lambda_1\cdot\vec \lambda_2$ & 4/3 & -8/3 \\
$ \vec \lambda_1\cdot\vec \lambda_3$ & -10/3 & -4/3\\
$ \vec \lambda_1\cdot\vec \lambda_4$ & -10/3 & -4/3\\
$ \vec \lambda_2\cdot\vec \lambda_3$ & -10/3 & -4/3\\
$ \vec \lambda_2\cdot\vec \lambda_4$ & -10/3 & -4/3\\
$ \vec \lambda_3\cdot\vec \lambda_4$ & 4/3 & -8/3\\
$\sum \vec \lambda_i\cdot\vec \lambda_j$& -32/3 &-32/3
\end{tabular}
\end{ruledtabular}
\end{table}
\begin{table}[b]
\caption{Ground and first radial excited meson states applied to fit the model parameters. The last column shows the deviation between the experimental and theoretical mean values, $D=100\cdot (M^{exp}-M^{cal})/M^{exp}$. $M^{exp}$ taken from PDG~\cite{PDG}.}
\label{tab:groundM}
\begin{ruledtabular}
\begin{tabular}{lccc}
Meson & $M^{exp}  {\rm (MeV)}$& $M^{cal} {\rm (MeV)}$  & D ($\%$)
\\
\hline
 $\Upsilon(1S)$ & 9460 & 9470 & -0.1
\\
 $\Upsilon(2S)$ & 10023 & 10070 &  -0.5
\\
 $\eta_b$ & 9399  & 9408 &  -0.1
\\
 $\eta_b(2S)$ & 9999 & 10008 &  -0.1
 \\
\hline
$J/\psi$ & 3097  &  3094 & 0.1
\\
 $\psi(2S)$ & 3686 & 3678 &  0.2
\\
$\psi(3S)$ & 4040 & 4053 &  -0.3
\\
 $\eta_c$ & 2984 & 2981 &  0.1
\\
 $\eta_c(2S)$ & 3638  & 3565 &  2.0
\\
\hline
 $B_s^*$ & 5415 & 5439 &  -0.4
\\
 $B_s^0$ & 5367 & 5310 &  1.1
\\
\hline
 $B^*$ & 5325 & 5367 &  -0.8
\\
 $B^0$ & 5279 & 5221 &  1.1
\\
\hline
 $D_s^*$ & 2112 & 2127 &  -0.7
\\
 $D_{s1}^*(2700)$ & 2708 & 2733 &  -0.9
\\
 $D_s$ & 1968 & 1981 &  -0.7
\\
\hline
 $D^*(2010)^0$ & 2010 & 2038 &  -1.4
\\
 $D^0$ & 1870 & 1878 & -0.4
\\
\hline
 $\phi(1020)$ & 1020 & 1029 &  -0.9
\\
 $\phi(1680)$ & 1680 & 1660 &  1.2
\\
\hline
 $\rho(770)$ & 770 & 779 &  -1.2
 \\
 $\rho(1450)$ & 1450 & 1436 &  1.0
 \\
\end{tabular}
\end{ruledtabular}
\end{table}

In line with the works discussed above, we propose $A_{ij}$ and $B_{ij}$ are massdependent coupling parameters in this work, taking the form
\begin{eqnarray}
A_{ij}= a+bm_{ij},\;\;B_{ij}=B_0 \sqrt{\frac{1}{m_{ij}}}.
\end{eqnarray}
with $m_{ij}$ being the reduced mass of $i$th and $j$th quarks, defined as $\;m_{ij}=\frac{2 m_i m_j}{m_i+m_j}$. $a$, $b$, and $B_0$ are constants. 

We propose the hyperfine interaction, $H_{hyp}$, has the same mass-dependence as the Coulomb-like interaction in Cornell potential, assuming that the hyperfine interaction and Coulomb-like interaction are from the same route of one gluon exchange. The massdependent hyperfine coefficient $C_{ij}$ takes the form, 
\begin{eqnarray}
 C_{ij} =C_0 \sqrt{\frac{1}{m_{ij}}}.
\end{eqnarray}
where $C_{0}$ is a constant. $\lambda^C_{i}$ and $\vec\sigma_i$ in Eq. (\ref{eqn::ham}) are the quark color operator and the spin operator respectively. The color matrix elements for the configurations in Eq. (\ref{eq:color}) are summarized in Table~\ref{tab:cme}.
\indent We calculate the mass spectra of light, strange, charmed and bottom mesons which are believed mainly $q\bar q$ states in the Hamiltonian in Eq. (\ref{eqn::ham}). The comparison of the theoretical results, as shown in Table~\ref{tab:groundM}, with the experimental data taken from Particle Data Group~\cite{PDG} leads to the four constituent quark masses and four model coupling parameters as follows.
\begin{eqnarray}\label{eq:nmo1}
&m_u = m_d = 420 \ {\rm MeV}\,, \quad m_s = 550 \ {\rm MeV}\,, \nonumber\\
&m_c = 1270 \ {\rm MeV}\,, \quad m_b = 4180 \ {\rm MeV}\,,  \nonumber\\
&a=67413 \ {\rm MeV^2}, \quad b=35 \ {\rm MeV}\,,  \nonumber\\  
&B_0=31.6635 \ {\rm MeV^{1/2}}, C_0=-188.765 \ {\rm MeV^{3/2}}\,.
\end{eqnarray}
We will apply the Hamiltonian in Eq. (\ref{eqn::ham}) with the predetermined as well as imported parameters to predict the mass of tetraquark states.

\section{\label{sec:PTM}Prediction of tetraquark masses of ground and lower radial excited states}
\indent The spin matrix elements for the spin combinations of charmoniumlike tetraquark states, as listed in Eq. (\ref{eq:spin}), and the spin combinations of fully-charm tetraquark states, as expressed in Eqs. (\ref{eq:spina6}) and (\ref{eq:spina3}), are summarized respectively in Tables~\ref{tab:smecl} and \ref{tab:smefc}.

The mass spectra of the ground and first radial excited charmoniumlike tetraquarks and the ground, first and second radial excited fully-charm tetraquarks are evaluated in the Hamiltonian in Eq. (\ref{eqn::ham}), where the complete bases defined in Sec. II are applied and the model parameters are predetermined in Sec. III. Listed in Tables \ref{tab:c1} and \ref{tab:fc1} are the theoretical results for charmoniumlike and fully-charm tetraquarks of various quark configurations, respectively.
\begin{table}[t]
\caption{Spin matrix elements of $qc\bar q\bar c$ charmoniumlike tetraquark states.}
\label{tab:smecl}
\begin{ruledtabular}
\begin{tabular}{lccccc}
$\hat{O}$ & $\psi^{S=0}_{0\otimes0}$ & $\psi^{S=0}_{1\otimes1}$ &  $\psi^{S=1}_{1\otimes0}$  & $\psi^{S=1}_{1\otimes1}$ &  $\psi^{S=2}_{1\otimes1}$ \\
\hline
$ \vec \sigma_1\cdot\vec \sigma_2$ & -3 & 1 & 1 & 1 & 1 \\
$ \vec \sigma_1\cdot\vec \sigma_3$ & 0 & -2 & 0 & -1 & 1 \\
$ \vec \sigma_1\cdot\vec \sigma_4$ & 0 & -2 & 0 & -1 & 1 \\
$ \vec \sigma_2\cdot\vec \sigma_3$ & 0 & -2 & 0 & -1 & 1 \\
$ \vec \sigma_2\cdot\vec \sigma_4$ & 0 & -2 & 0 & -1 & 1 \\
$ \vec \sigma_3\cdot\vec \sigma_4$ & -3 & 1 & -3 & 1 & 1\\
$\sum \vec \sigma_i\cdot\vec \sigma_j$ & -6 & -6 & -2 & -2 & 6
\end{tabular}
\end{ruledtabular}
\end{table}
\begin{table}[b]
\caption{Spin matrix elements of $cc\bar c \bar c$ fully-charm tetraquark states.}
\label{tab:smefc}
\begin{ruledtabular}
\begin{tabular}{lcccc}
$\hat{O}$ & $\psi_{(6\otimes\bar 6)(0\otimes0)}^{C, S=0}$ & $\psi_{(\bar 3\otimes3)(1\otimes1)}^{C, S=0}$ & $\psi_{(\bar 3\otimes3)(1\otimes1)}^{C,S=1}$ &  $\psi_{(\bar 3\otimes3)(1\otimes1)}^{C,S=2}$ \\
\hline
$ \vec \sigma_1\cdot\vec \sigma_2$ & -3 & 1 & 1 & 1\\
$ \vec \sigma_1\cdot\vec \sigma_3$ & 0 & -2 & -1 & 1 \\
$ \vec \sigma_1\cdot\vec \sigma_4$ & 0 & -2 & -1 & 1 \\
$ \vec \sigma_2\cdot\vec \sigma_3$ & 0 & -2 & -1 & 1 \\
$ \vec \sigma_2\cdot\vec \sigma_4$ & 0 & -2 & -1 & 1 \\
$ \vec \sigma_3\cdot\vec \sigma_4$ & -3 & 1 & 1 & 1\\
$\sum \vec \sigma_i\cdot\vec \sigma_j$ & -6 & -6 & -2 & 6
\end{tabular}
\end{ruledtabular}
\end{table}
\begin{table}[t]
\caption{\label{tab:c1}Ground and first radial excited charmoniumlike tetraquark masses (MeV).}
\begin{ruledtabular}
\begin{tabular}{lccccc}
$qc\bar q\bar c$ configurations & $I^GJ^{PC}$ &  M(1S) & M(2S)
\\
\hline
$\Psi^{cs}_{(6\otimes\bar 6)[(qc)^{s=0}\otimes(\bar q\bar c)^{s=0})]^{S=0}}$ & $0^+0^{++}/1^-0^{++}$ & 4202 & 4566
\\
 $\Psi^{cs}_{(\bar 3\otimes3)[(qc)^{s=0}\otimes(\bar q\bar c)^{s=0})]^{S=0}}$ & $0^+0^{++}/1^-0^{++}$ & 4033 & 4434
\\
\hline
 $\Psi^{cs}_{(6\otimes\bar 6)[(qc)^{s=1}\otimes(\bar q\bar c)^{s=1})]^{S=0}}$ & $0^+0^{++}/1^-0^{++}$ & 3925 & 4289
\\
 $\Psi^{cs}_{(\bar 3\otimes3)[(qc)^{s=1}\otimes(\bar q\bar c)^{s=1})]^{S=0}}$ & $0^+0^{++}/1^-0^{++}$ & 4114 & 4516
\\
\hline
 $\Psi^{cs}_{(6\otimes\bar 6)[(qc)^{s=1}\otimes(\bar q\bar c)^{s=0})]^{S=1}}$ & $0^-1^{+-}/1^+1^{+-}$ & 4162 & 4526
\\
 $\Psi^{cs}_{(\bar 3\otimes3)[(qc)^{s=1}\otimes(\bar q\bar c)^{s=0})]^{S=1}}$ & $0^-1^{+-}/1^+1^{+-}$ & 4113 & 4514
\\
\hline
 $\Psi^{cs}_{(6\otimes\bar 6)[(qc)^{s=1}\otimes(\bar q\bar c)^{s=1})]^{S=1}}$ & $0^-1^{+-}/1^+1^{+-}$ & 4024 & 4388
\\
 $\Psi^{cs}_{(\bar 3\otimes3)[(qc)^{s=1}\otimes(\bar q\bar c)^{s=1})]^{S=1}}$ & $0^-1^{+-}/1^+1^{+-}$ & 4154 & 4555
\\
\hline
 $\Psi^{cs}_{(6\otimes\bar 6)[(qc)^{s=1}\otimes(\bar q\bar c)^{s=1})]^{S=2}})$ & $0^+2^{++}/1^-2^{++}$ & 4221 & 4584
 \\
 $\Psi^{cs}_{(\bar 3\otimes3)[(qc)^{s=1}\otimes(\bar q\bar c)^{s=1})]^{S=2}}$ & $0^+2^{++}/1^-2^{++}$ & 4233 & 4634
\end{tabular}
\end{ruledtabular}
\end{table}
\begin{table}[t]
\caption{\label{tab:fc1} Ground, first and second radial excited fully-charm tetraquark masses (MeV).}
\begin{ruledtabular}
\begin{tabular}{lccccc}
$cc\bar c\bar c$ configurations & $I^GJ^{PC}$ &  M(1S) &M(2S) & M(3S)
\\
\hline
 $\Psi^{cs}_{(6\otimes\bar 6)[(cc)^{s=0}\otimes(\bar c\bar c)^{s=0})]^{S=0}}$ & $0^+0^{++}$ & 6514 & 6840 & 7098
\\
 $\Psi^{cs}_{(\bar 3\otimes3)[(cc)^{s=0}\otimes(\bar c\bar c)^{s=0})]^{S=0}}$ & $0^+0^{++}$
 & 6466 & 6883 & 7225
\\
\hline
 $\Psi^{cs}_{(\bar 3\otimes3)[(cc)^{s=1}\otimes(\bar c\bar c)^{s=1})]^{S=1}}$ & $0^-1^{+-}$
 & 6494 & 6911 & 7253
\\
\hline
 $\Psi^{cs}_{(\bar 3\otimes3)[(cc)^{s=1}\otimes(\bar c\bar c)^{s=1})]^{S=2}}$ & $0^+2^{++}$
 & 6551 & 6968 & 7310
\end{tabular}
\end{ruledtabular}
\end{table}

\section{\label{sec:DIS}Discussion}

\begin{table*}[t]
\caption{\label{tab:A} Tentative assignments of ground and first radial excited charmoniumlike tetraquark states}
\begin{ruledtabular}
\begin{tabular}{ccccccccccc}
$qc\bar q\bar c$ & \multicolumn{2}{c}{$\psi^{S=0}_{0_s\otimes0_s}$} & \multicolumn{2}{c}{$\psi^{S=0}_{1_s\otimes1_s}$} & \multicolumn{2}{c}{$\psi^{S=1}_{1_s\otimes0_s}$} & \multicolumn{2}{c}{$\psi^{S=1}_{1_s\otimes1_s}$} & \multicolumn{2}{c}{$\psi^{S=2}_{1_s\otimes1_s}$}
\\
configurations & \multicolumn{2}{c}{$0^+0^{++}/1^-0^{++}$} & \multicolumn{2}{c}{$0^+0^{++}/1^-0^{++}$} & \multicolumn{2}{c}{$0^-1^{+-}/1^+1^{+-}$} & \multicolumn{2}{c}{$0^-1^{+-}/1^+1^{+-}$} & \multicolumn{2}{c}{$0^+2^{++}/1^-2^{++}$}
\\
& Ours & Exp. & Ours & Exp. & Ours & Exp. & Ours & Exp. & Ours & Exp.
\\
\hline
$\Psi^{c}_{6_c\otimes\bar 6_c}(1S)$ & 4202 & $Z_c(4250)$ & 3925 & $X(3915)$ & 4162 & $Z_c(4200)$ & 4024 & $Z_c(4020)$/$Z_c(4055)$ & 4221 & --
\\
$\Psi^{c}_{6_c\otimes\bar 6_c}(2S)$ & 4566 & -- & 4289 & $X(4350)$ & 4526 & $Z_c(4430)$ & 4388 & -- & 4584 & --
\\
\hline
$\Psi^{c}_{\bar 3_c\otimes3_c}(1S)$ & 4033 & $Z_c(4050)$ & 4114 & $Z_c(4100)$ & 4113 & $X(4160)$ & 4154 & $X(4160)$ & 4233 & --
\\
$\Psi^{c}_{\bar 3_c\otimes3_c}(2S)$ & 4434 & -- & 4516 & -- & 4514 & -- & 4555 & -- & 4634 & --
\end{tabular}
\end{ruledtabular}
\end{table*}

As a number of exotic particles have been discovered, recent years can be called  a revolutionary period in the field of hadron physics. Among the exotic particles, charmoniumlike charged mesons may be especially interesting since they have a charmoniumlike mass but are electrically charged \cite{Albuquerque:2018jkn}. Those charged charmoniumlike states go beyond the conventional $c\bar c$-meson picture and are likely tetraquark states $c\bar cu\bar d$ due to carrying one charge. We have listed these states in Table \ref{tab:particle}, denoting all of them with X, Y and $Z_c$, which are more convenient for referring to other works than the latest naming scheme of the Particle Data Group~\cite{PDG}. We shall use X and $Z_c$ , throughout this discussion, to refer to neutral states and charged states with hidden charm respectively.

\subsection{charmoniumlike tetraquark states}

$Z_c(4050)$ and $Z_c(4250)$ have been studied as tetraquarks in different models. $Z_c(4050)$ was investigated as a cluster of $Q\bar q$ and $\bar Qq$ in a Cornell-like potential with some residual color forces that bind the two clusters \cite{Patel:2014vua}, which results in two states with masses 4046 and 4054 MeV with quantum numbers $J^{PC} = 2^{+-}$ and $J^{PC} =3^{++}$ respectively. Both states are associated with $Z_c(4050)$.
In the color flux-tube model, a conclusion was made that $Z_c(4050)$ has a tetraquark $(cu)(\bar c\bar d)$ nature, with spin-parity $J^P = 1^-$, while $Z_c(4250)$ could be interpreted as a $(cu)(\bar c\bar d)$ tetraquark with $J^P = 1^+$\cite{Deng:2015lca}. No tetraquark candidate was found for $Z_c(4050)$ in a relativistic quark model, but $Z_c(4250)$ could be interpreted as a tetraquark state \cite{Ebert:2008kb}.

Considering that both $Z_c(4050)$ and $Z_c(4250)$ were observed in the process $\bar B^0 \to K^-\pi^+\chi_{c1}$ and their tentative quantum numbers \cite{Mizuk:2008me}, the present predictions support assigning the $Z_c(4050)$ and $Z_c(4250)$ to be the ground states with $J^{PC} = 0^{++}$ of the $(\bar 3_c \otimes 3_c)(0_s\otimes0_s)_{S=0}$ and $(6_c \otimes \bar 6_c)(0_s\otimes0_s)_{S=0}$ configurations, respectively.

The tetraquark nature of $Z_c(4200)$ has been studied in various model calculations as well. A $(cu)(\bar c\bar d)$ state was predicted, with the quantum numbers $n(^{2S+1}L_J)=1(^3D_1)$ and spin-parity $1^+$, which is associated with $Z_c(4200)$, in a model treating quark-quark interactions through one gluon exchange, one boson exchange and $\sigma$ exchange\cite{Deng:2017xlb}. The study in a light-front holographic QCD model preferred that $Z_c(4200)$ having a generic dilaton profile\cite{Guo:2016uaf}. Using a formalism based on color magnetic interactions, $Z_c(4200)$ was described as an axial vector tetraquark state\cite{Zhao:2014qva}.

After its discovery, the $Z^+_c(3900)$ was identified as the predicted $X^+$\cite{Faccini:2013lda}, and $Z^+(4430)$ was identified as the first radial excitation of $Z^+_c(3900)$\cite{Maiani:2014aja}. $Z_c^+(4430)$ was also interpreted as the first radial excitation (2S) of a charged diquark-antidiquark $(cu)(\bar c\bar d)$ tetraquark state in Refs\cite{Goerke:2016hxf, Ebert:2008kb, Patel:2014vua, Wang:2014vha, Agaev:2017tzv}.

Since both $Z_c(4200)$ and $Z_c(4430)$ were observed in the process $\bar B^0\to K^-\pi^+J/\psi $ \cite{Chilikin:2014bkk} and their decay widths are in the same order which is much larger than the $Z^+_c(3900)$ one, one may naturally pair the $Z_c(4200)$ and $Z_c(4430)$ together. Therefore, we may assign the $Z_c(4200)$ and $Z_c(4430)$ to be the ground and first radial excited states, with $J^{PC} = 1^{+-}$, of the $(6_c \otimes \bar 6_c)(1_s\otimes0_s)_{S=1}$ configuration, respectively.

In the color flux tube model with a multibody confinement potential, tetraquark states were studied in the diquark-antidiquark configuration, and it was found that the nearest state to $Z^+_c( 4025) $ obtained by the model is the one with quantum number $J^P=2^+$\cite{Deng:2015lca, Deng:2014gqa}. However, more works support the $1^{+-}$ quantum numbers.
In the framework of nonrelativistic quark model and applying a Cornell-type potential, a molecularlike four-quark state of $ Q\bar q-\bar Qq $ with $ J^{PC}=1^{+-} $ was predicted around 4036 MeV, among others with similar masses but other quantum numbers \cite{Patel:2014vua}. This state was identified with $Z_c(4025)$. $Z_c(4025)$ and $Z_c(4020)$ are named as $X(4020)$ nowadays in PDG~\cite{PDG}. Except for the parity, other quantum numbers of the $X(4020)$ are not well determined, but all the experimental analyses from BESIII assumed s-wave productions and the quantum number assignment $J^{PC}=1^{+-}$\cite{Ablikim:2013wzq, Ablikim:2013emm, Ablikim:2014dxl, Ablikim:2015vvn}.

\begin{figure}[b]
\centering
\includegraphics[width=0.4  \textwidth]{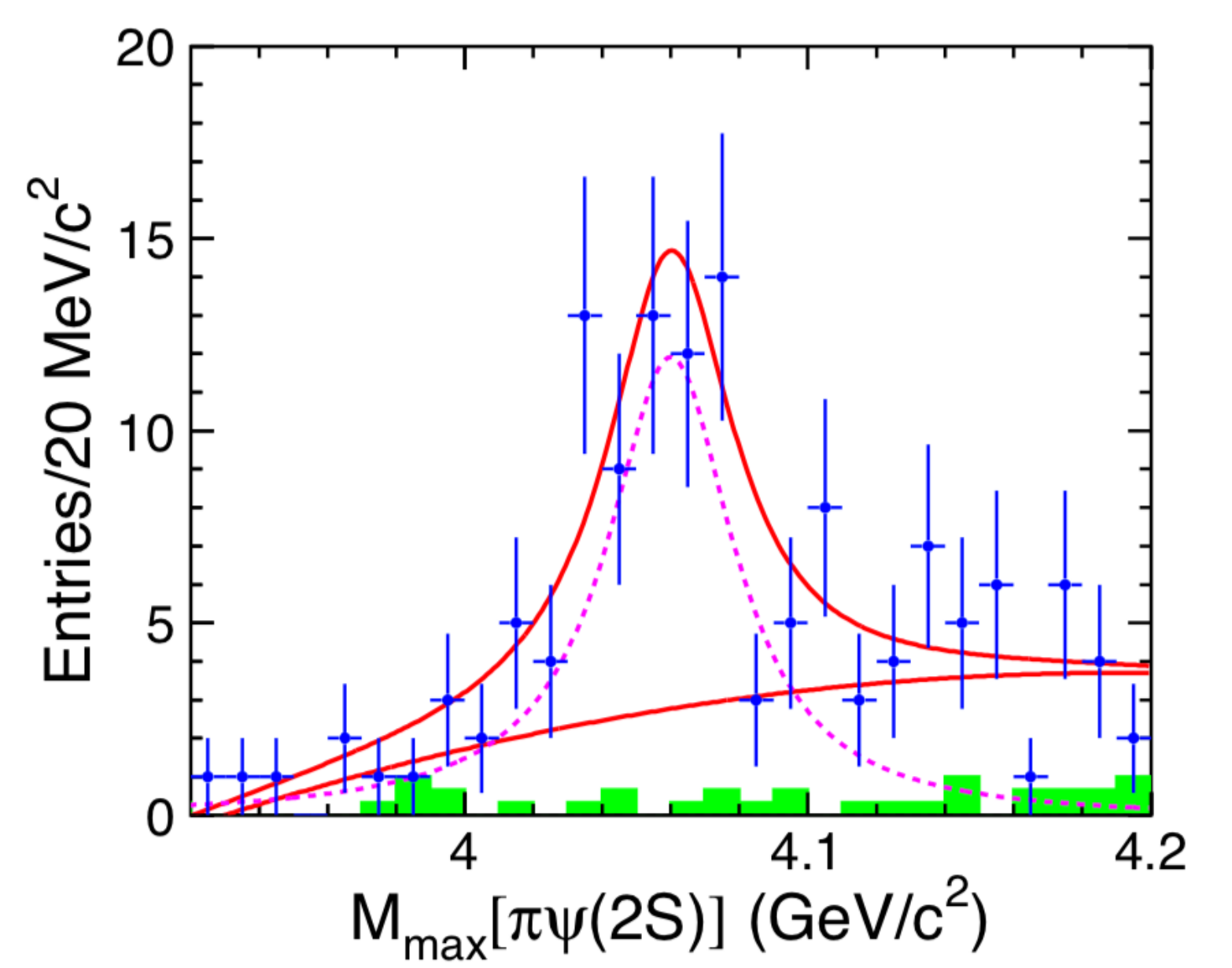}
\caption{\label{fig:belle} The distributions of $M_{max}[\pi^{\pm}\psi(2S)]$ from $Y(4360)$ decays \cite{Wang:2014hta}.}
\end{figure}

\begin{figure}[b]
\centering
\includegraphics[width=0.4  \textwidth]{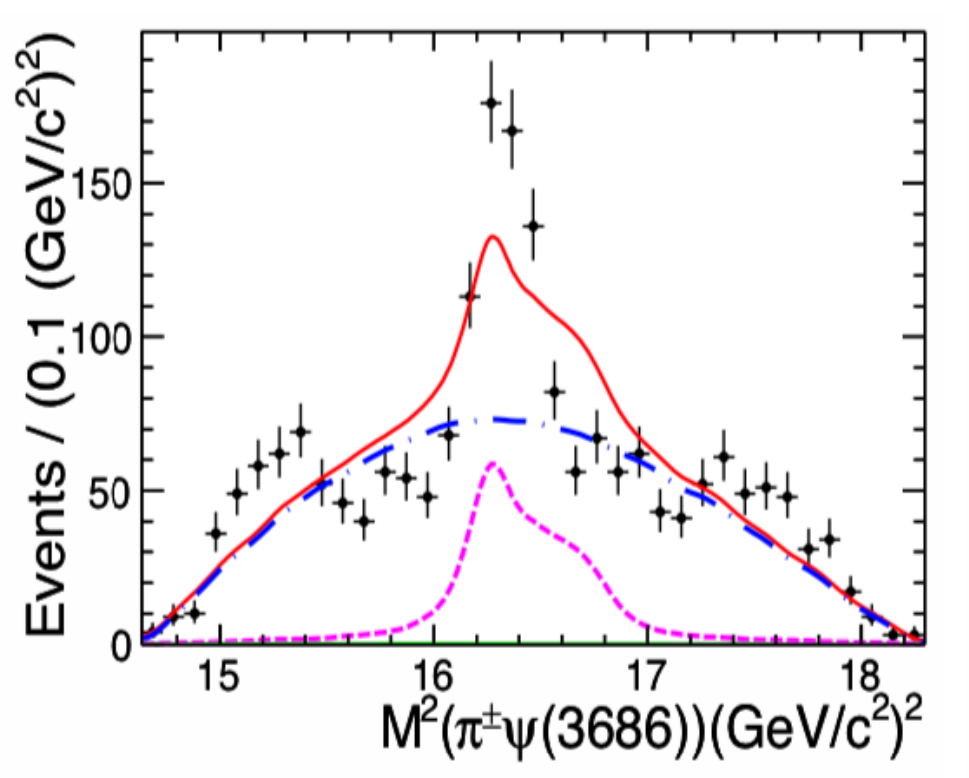}
\caption{\label{fig:bes3} The distributions of $M[\pi^{\pm}\psi(2S)]$ at $\sqrt s =4.416$ GeV (two entries per event) from BESIII \cite{Ablikim:2017oaf}.}
\end{figure}

Belle updated the measurement of $e^+e^-\to\pi^+\pi^-\psi(2S)$ via ISR using the 980 $fb^{-1}$ full data sample\cite{Wang:2014hta}. Fig. \ref{fig:belle} showed the $M_{max}[\pi^{\pm}(2S)]$ distribution, the maximum of $M[\pi^-\psi(2S)]$ and $M[\pi^+\psi(2S)]$, in $Y(4360)$ decays $(4.0$ GeV$<M_{\pi^+\pi^-\psi(2S)}<4.5$ GeV$)$, where an excess evidence at around 4.05 GeV can be seen. The fit yields a mass of $(4054\pm3\pm1)$ MeV and a width of $(45\pm11\pm6)$ MeV. The statistical significance of the signal is $3.5\sigma$ with systematic uncertainties included.

BESIII studied the process $e^+e^-\to\pi^+\pi^-h_c$ at c.m. energies from $3.90$ to $4.42$ GeV\cite{Ablikim:2013wzq}. In the $\pi^{\pm}h_c$ mass spectrum, a distinct structure, referred to as $Z_c(4020)$, was observed at $4.02$ GeV.  The $Z_c(4020)$ carries an electric charge and couples to charmonium. A fit to the $\pi^{\pm}h_c$ invariant mass spectrum, neglecting possible interferences, results in a mass of $(4022.9\pm0.8\pm2.7)$ MeV and a width of $(7.9\pm2.7\pm2.6)$ MeV for the $Z_c(4020)$, where the first errors are statistical and the second systematic.
Later, BESIII studied the process $e^+e^-\to\pi^+\pi^-\psi(2S)$ using $5.1 fb^{-1}$ of data at c.m. energies from 4.0 to 4.6 GeV\cite{Ablikim:2017oaf}. For data at $\sqrt s =4.416$ GeV, a prominent narrow structure was observed around 4030 MeV in the $M[\pi^{\pm}\psi(2S)]$ spectrum, as shown in Fig.\ref{fig:bes3}. The fit yields a mass of $M=(4032.1\pm2.4)$ MeV and a width of $\Gamma=(26.1\pm5.3)$ MeV for the intermediate state with a much higher significance than Belle of $9.2\sigma$.

The authors of Ref.\cite{Bondar2018} reported their preliminary PWA results on $e^+e^-\to\pi^+\pi^-\psi(2S)$ at the Charm 2018 meeting by using BESIII published results\cite{Ablikim:2017oaf}. The fit quality is much improved. It was found that the structure can be described well with a charged state with a mass of $(4019.0\pm1.9)$ MeV and a width of $(29\pm4)$ MeV, or the $Z_c(4020)$ state observed in the $\pi^+\pi^-h_c$ final state\cite{Ablikim:2013wzq}. If such PWA results are confirmed in the future, one may argue that $Z_c(4020)$ may have another decay mode $Z_c(4020)\to\pi^+\psi(2S)$ and furthermore $Z_c(4020)$ and $Z_c(4055)$ could be the same state. Thus, both $\pi^+\pi^-\psi(2S)$ and $\pi^+\pi^-h_c$ final states need to be further investigated to understand the intermediate structures. At this moment, we may just assign either $Z_c(4020)$ or $Z_c(4055)$, if they are not the same particle, to be the $(6_c \otimes \bar 6_c)(1_s\otimes1_s)_{s=1}$ configuration tetraquark ground states with $J^{PC} = 1^{+-}$.

Up to now, there have been only few theoretical calculations for $Z_c(4100)$. A simple chromomagnetic model was employed to study the mass splitting among tetraquark states, including $Z_c(4100)$\cite{Wu:2018xdi}. The model is based on the description that the mass splitting among hadron states, with the same quark content, are mainly due to the chromomagnetic interaction term in the one-gluon-exchange potential. Based on these findings, it was concluded that $Z_c(4100)$ seems to be a $0^{++}$ $(cq)(\bar c \bar q)$ tetraquark state. In our assignments, $Z_c(4100)$ is assigned to be the $(\bar3_c \otimes 3_c)(1_s\otimes1_s)_{S=0}$ configuration tetraquark ground state with $J^{PC} = 0^{++}$.

Considering that the $X(3915)$ and $X(4350)$ have similar decay patterns, that is, the $X(3915)$ state decays mainly to $J/\psi \omega$ and the $X(4350)$ state was observed only in the process $\gamma \gamma \to \phi J/\psi $, and that their decay widths were narrow and in the same order, we may assign $X(3915)$ and $X(4350)$ together to be ground and first radial excited states, with $J^{PC}=0^{++}$, of the $(6_c \otimes \bar 6_c)(1_s\otimes1_s)_{S=0}$ configuration, respectively.

There is no room to accommodate the $Z_c(3900)$ in the scenario of tetraquark states in the present model. As discussed above, there are two possible explanations for the $Z^+_c(3900)$ structure: a charged diquark-antidiquark $(cu)(\bar c\bar d)$ state, and a $D\bar D^*$ molecular state. Concerning the molecular configuration, there are many calculations that could not accommodate $Z^+_c(3900)$ as a $J^P = 1^+$ $D\bar D^*$ molecule, including lattice QCD calculations \cite{Zhao:2014gqa, He:2014nya,Prelovsek:2013xba, Chen:2014afa}. In Refs.\cite{Wang:2013daa, Aceti:2014uea, Ke:2016owt}, however, a $D\bar D^*$ molecular state is derived, compatible with $Z^+_c(3900)$.  $Z_c(3900)$ may be axial vector moleculelike state with $J^{PC} = 1^{+-}$.

Except the assignments discussed above, $X(4160)$ is also tentatively assigned to be the $(\bar 3_c \otimes 3_c)(1_s\otimes0_s)_{S=1}$ and $(\bar 3_c \otimes 3_c)(1_s\otimes1_s)_{S=1}$ configuration tetraquark ground states with $J^{PC} = 1^{+-}$ according to the mass matching.

Here, in Table \ref{tab:A}, almost all the ground charmoniumlike tetraquark states predicted in the work have been tentatively matched with experimental data. Seven in eight charged charmoniumlike tetraquark states observed by experimental collaborations have been assigned in the work except $Z^+_c(3900)$.  More experimental data and theoretical works are essential to make unambiguous assignments. In addition, there is no room of the ground and first excited states to accommodate the $X(3860)$, $X(3940)$, $X(4500)$, and $X(4700)$ in the scenario of tetraquark states in the present model.

\subsection{Fully-charm tetraquark states}

\begin{table*}[tb]
\caption{\label{tab:fc2} Present predictions of ground state fully-charm tetraquark masses (MeV) compared with others.}
\begin{ruledtabular}
\begin{tabular}{lccccccccccccc}
$cc{\bar c\bar c}$ configurations & $J^{PC}$ &  Ours & \cite {Yang:2020rih}& \cite{Wang:2019rdo} &\cite{Liu:2019zuc} & \cite{ Lloyd:2003yc} & \cite{Ader:1981db} & \cite{Chen:2020xwe}  & \cite{Barnea:2006sd} & \cite{Berezhnoy:2011xn} & \cite{Wang:2017jtz}\cite{Wang:2018poa} & \cite{Debastiani:2017msn}
\\
\hline
$\Psi^{cs}_{(6_c\otimes\bar 6_c)(0_s\otimes0_s)^{S=0}}$ & $0^{++}$ & 6514 & 6404 & 6383-6421& 6518 &  6695 & 6383 & 6440-6820 & \multirow{2}{*}{6038-6115} & \multirow{2}{*}{5966} & \multirow{2}{*}{5990} & \multirow{2}{*}{5969}
\\
$\Psi^{cs}_{(\bar 3_c\otimes3_c)(1_s\otimes1_s)^{S=0}}$ & $0^{++}$ & 6466 & 6421 & 6420-6436 & 6487 & 6477  & 6437 & 6460-6470   &  &  &  &
\\
\hline
$\Psi^{cs}_{(\bar 3_c\otimes3_c)(1_s\otimes1_s)^{S=1}}$ & $1^{+-}$ & 6494 & 6439 & 6425-6450 & 6500 & 6528  & 6437 & 6370-6510 & 6101-6176 & 6051 & 6050 &6021
\\
\hline
$\Psi^{cs}_{(\bar 3_c\otimes3_c)(1_s\otimes1_s)^{S=2}}$ & $2^{++}$ & 6551 & 6472 & 6432-6479 & 6524 & 6573 & 6437  & 6370-6510 & 6172-6216 & 6223 & 6090 &6115
\\
\end{tabular}
\end{ruledtabular}
\end{table*}

\begin{figure}[tb]
\centering
\includegraphics[width=0.5  \textwidth]{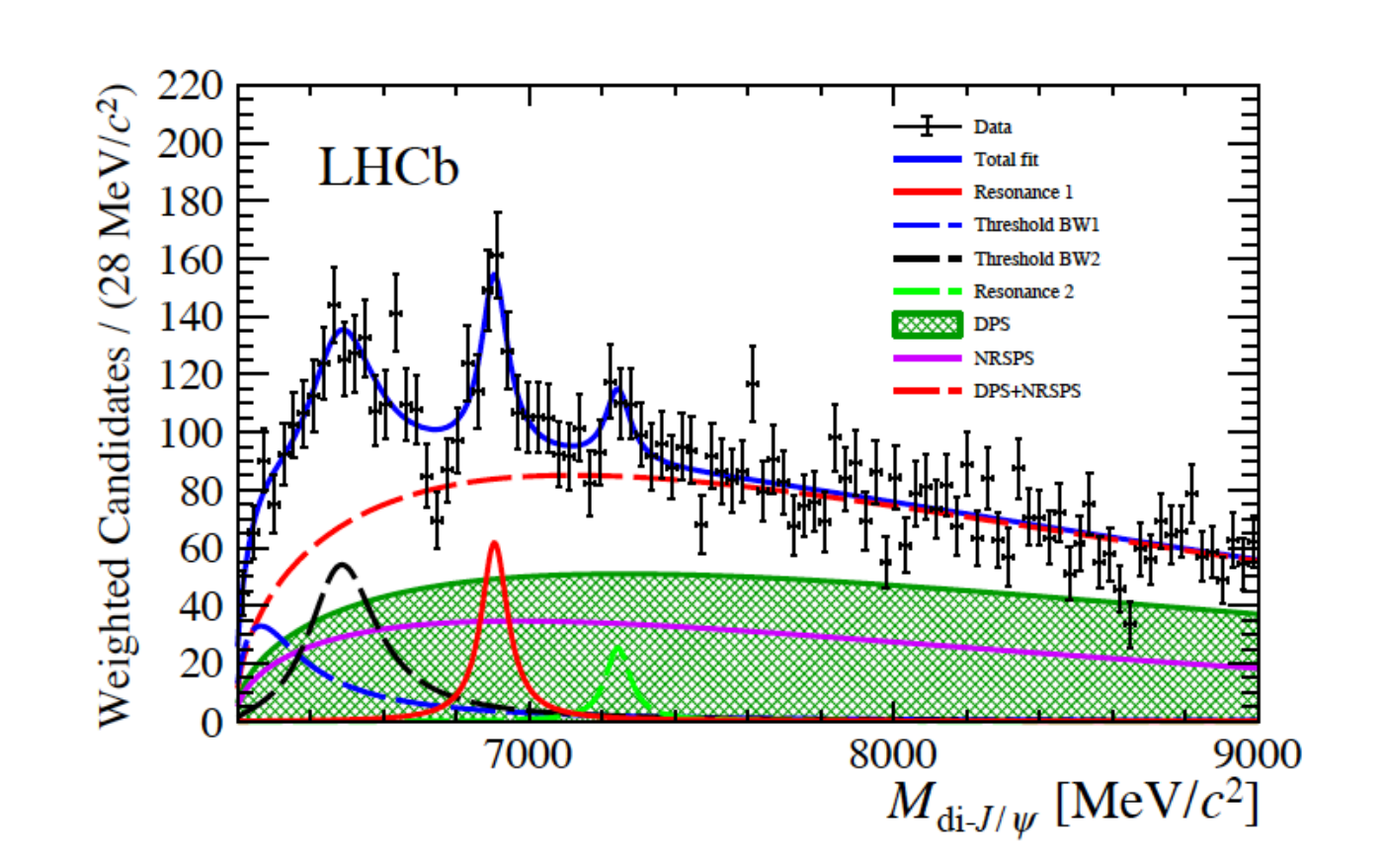}
\caption{\label{fig:fc}Invariant mass spectra of weighted di-$J/\psi$ candidates. This was adapted from figure S3 in \cite{Aaij:2020fnh}.}
\end{figure}

The LHCb collaboration has recently presented evidence for the observation of at least one resonance in the $J/\psi$-pair spectrum at about 6900 MeV\cite{Aaij:2020fnh}. The data also revealed a broader structure centered around 6500 MeV, as shown in Figure \ref{fig:fc}. Such states are naturally assigned the valence-quark content $c\bar cc\bar c$, making them the first all-heavy multiquark exotic candidates claimed to date in the experimental literature. The measured mass and width of the resonance around 6900 MeV are
\begin{flalign}\label{mdm1}
M[X(6900)]=6905\pm11\pm7 \nonumber \\
\Gamma[X(6900)]=80\pm19\pm33
\end{flalign}
or, using a second fitting model
\begin{flalign}\label{mdm2}
M[X(6900)]=6886\pm11\pm11 \nonumber \\
\Gamma[X(6900)]=168\pm33\pm69
\end{flalign}

Our present theoretical predictions, as shown in Table~\ref{tab:fc1}, support assigning the $X(6900)$ to be the first radial excited fully-charm tetraquark state, with $J^{PC} = 1^{+-}$, in the $\bar 3_c \otimes 3_c$ configuration. The ground and second radial excited states of fully-charm tetraquarks are around 6494 MeV and 7253 MeV respectively, with $J^{PC} = 1^{+-}$, in the $\bar 3_c \otimes 3_c$ configuration, which is compatible with the experimental data in Figure \ref{fig:fc}.

We collect our numerical results and some typical results of other works in Table \ref {tab:fc2} for comparison. Including our results, the predictions of nonrelativistic quark models are roughly compatible \cite{Yang:2020rih, Wang:2019rdo, Liu:2019zuc, Ader:1981db, Lloyd:2003yc}, where both confining and OGE Coulomb-like potentials are considered.
One work in QCD sum rules also gives similar results \cite{Chen:2020xwe}.
Refs.\cite{Barnea:2006sd, Berezhnoy:2011xn, Wang:2017jtz, Wang:2018poa, Debastiani:2017msn} give smaller masses, without considering color configurations.

\section{Summary}\label{sec:SUM}
We have evaluated the masses of ground and first radial excited charmoniumlike tetraquark states and of ground and first and second radial excited states of the fully-charm tetraquark states, with all model parameters predetermined by comparing the theoretical and experimental masses of light, charmed and bottom mesons. 

We have made a tentative matching between the predicted ground and first radial excited charmoniumlike tetraquark states and the X and Z particles. The present work predicts some charmoniumlike tetraquark states which can not be matched with observed particles. One may suggest experimental searchings in the processes $e^+e^-\to \pi^{\mp}(\pi^{\pm} h_c),\;\pi^{\mp}(\pi^{\pm}\psi(2S))$ for higher mass resonances, probably the first radial excited states of $Z_c(4020)$, and $Z_c(4055)$.

The work suggests that the $X(6900)$ observed by LHCb is likely the first radial excited fully-charm tetraquark state, with $J^{PC} = 1^{+-}$, in the $\bar 3_c \otimes 3_c$ configuration. The ground and second radial excited states of fully-charm tetraquarks, with $J^{PC} = 1^{+-}$, in the $\bar 3_c \otimes 3_c$ configuration are around 6494 MeV and 7253 MeV, respectively.

\begin{acknowledgments}
This work is supported by Suranaree University of Technology (SUT). Z. Zhao acknowledges support from the Institute of Research and development, Suranaree University of Technology OROG Ph.D. scholarship under the contract No. 62/2559. K. Xu, A. Kaewsnod, A. Limphirat, and Y. Yan acknowledge support from SUT. X.Y. Liu acknowledges support from Young Science Foundation from the Education Department of Liaoning Province, China (Project No. LQ2019009).\end{acknowledgments}

\appendix

\section{Construction of tetraquark wave functions}\label{sec:AP1}

The construction of tetraquark states follows the rule that a tetraquark state must be a color singlet and the $qq$ and $\bar q\bar q$ clusters wave function should be antisymmetric under any permutation between identical quarks. Requiring the tetraquark to be a color singlet demands that the color part of the $qq$ and $\bar q\bar q$ must form a $SU_c(3)$ $[222]_1$ singlet state, and hence two sets of color configurations, $[2](q_1c_2)\otimes[22](\bar q_3\bar c_4)$ and $ [11](q_1c_2)\otimes[211](\bar q_3\bar c_4)$, taking charmoniumlike tetraquark states as an
example. We may take charmoniumlike tetraquark states as an example. The order of quarks is defined as $q_1c_2\bar q_3\bar c_4$.
\begin{flalign}\label{eq:gcc}
&[2](q_1c_2)\otimes[22](\bar q_3\bar c_4) \ and \ [11](q_1c_2)\otimes[211](\bar q_3\bar c_4),
\end{flalign}
where denote color configurations sextet-antisextet ($6_c \otimes \bar 6_c$) and triplet-antitriplet ($\bar 3_c \otimes 3_c$).

The color wave function of each tetraquark color configuration may be written in the general form
\begin{flalign}\label{eq:colorag}
&\psi^{q_1c_2\bar q_3\bar c_4}_{[2]_{6}^c{{[22]}^c_{\bar 6}}}=\frac 1{\sqrt 6}\sum_{i=1}^{6}\psi_{[2]_{6}^ci}^{q_1c_2}\psi_{{[22]}^c_{\bar 6}i}^{\bar q_3 \bar c_4},
\nonumber \\
&\psi^{q_1c_2\bar q_3\bar c_4}_{[11]_{\bar 3}^c{{[211]}^c_{3}}}=\frac 1{\sqrt 3}\sum_{i=1}^{3}\psi_{[11]_{\bar 3}^ci}^{q_1c_2}\psi_{{[211]}^c_{3}i}^{\bar q_3 \bar c_4},
\end{flalign}

The explicit color wave functions without subscripts are using the quark order $q_1c_2\bar q_3\bar c_4$, listed as follows
\begin{flalign}\label{eq:colora6}
\psi^{q_1c_2\bar q_3\bar c_4}_{[2]_{6}^c{{[22]}^c_{\bar 6}}}&=\frac{1}{\sqrt6}[R_1R_2\bar R_3\bar R_4+G_1G_2\bar G_3\bar G_4+B_1B_2\bar B_3\bar B_4\nonumber \\
&+\frac12(R_1G_2+G_1R_2)(\bar R_3\bar G_4+\bar G_3 \bar R_4)\nonumber \\
&+\frac12(B_1R_2+R_1B_2)(\bar B_3\bar R_4+\bar R_3 \bar B_4)\nonumber \\
&+\frac12(G_1B_2+B_1G_2)(\bar G_3\bar B_4+\bar B_3 \bar G_4)]\nonumber \\
&=\frac{1}{\sqrt6}[RR\bar R\bar R+GG\bar G\bar G+BB\bar B\bar B\nonumber \\
&+\frac12(RG\bar R\bar G+GR\bar R\bar G+RG\bar G\bar R+GR\bar G \bar R)\nonumber \\
&+\frac12(BR\bar B\bar R+RB\bar B\bar R+BR\bar R \bar B+RB\bar R \bar B)\nonumber \\
&+\frac12(GB\bar G\bar B+BG\bar G\bar B+GB\bar B \bar G+BG\bar B \bar G)]
\end{flalign}

\begin{flalign}\label{eq:colora3}
\psi^{q_1c_2\bar q_3\bar c_4}_{[11]_{\bar 3}^c{{[211]}^c_{3}}}&=\frac{1}{\sqrt3}[\frac12(R_1G_2-G_1R_2)(\bar R_3\bar G_4-\bar G_3 \bar R_4)\nonumber \\
&+\frac12(B_1R_2-R_1B_2)(\bar B_3\bar R_4-\bar R_3 \bar B_4)\nonumber \\
&+\frac12(G_1B_2-B_1G_2)(\bar G_3\bar B_4-\bar B_3 \bar G_4)]\nonumber \\
&=\frac{1}{\sqrt3}[\frac12(RG\bar R\bar G-GR\bar R\bar G-RG\bar G\bar R+GR\bar G \bar R)\nonumber \\
&+\frac12(BR\bar B\bar R-RB\bar B\bar R-BR\bar R \bar B+RB\bar R \bar B)\nonumber \\
&+\frac12(GB\bar G\bar B-BG\bar G\bar B-GB\bar B \bar G+BG\bar B \bar G)]
\end{flalign}

The explicit spin wave functions $\psi^{S(qc\bar q\bar c)}_{(S(qc)\otimes S(\bar q\bar c))}$ of $qc\bar q\bar c$ tetraquark states are listed as follows

\begin{flalign}\label{eq:spincl}
\psi^{S=2}_{(1\otimes1)}&=\uparrow \uparrow \bar \uparrow \bar \uparrow, \nonumber \\
\psi^{S=1}_{(1\otimes1)}&=\frac 12(\uparrow \uparrow \bar \uparrow \bar \downarrow+\uparrow \uparrow \bar \downarrow \bar \uparrow-\uparrow \downarrow \bar \uparrow \bar \uparrow-\downarrow \uparrow \bar \uparrow \bar \uparrow), \nonumber \\
\psi^{S=1}_{(1\otimes0)}&=\frac 1{\sqrt2}(\uparrow \uparrow \bar \uparrow \bar \downarrow-\uparrow \uparrow \bar \downarrow \bar \uparrow), \nonumber \\
\psi^{S=0}_{(1\otimes1)}&=\frac 1{\sqrt 3}[\uparrow \uparrow \bar \downarrow \bar \downarrow-\frac 12(\uparrow \downarrow \bar \uparrow \bar \downarrow+\uparrow \downarrow \bar \downarrow \bar \uparrow+\downarrow \uparrow \bar \uparrow \bar \downarrow+\downarrow \uparrow \bar \downarrow \bar \uparrow) \nonumber \\
& +\downarrow \downarrow \bar \uparrow \bar \uparrow], \nonumber \\
\psi^{S=0}_{(0\otimes0)}&=\frac 12(\uparrow \downarrow \bar \uparrow \bar \downarrow-\uparrow \downarrow \bar \downarrow \bar \uparrow-\downarrow \uparrow \bar \uparrow \bar \downarrow+\downarrow \uparrow \bar \downarrow \bar \uparrow)
\end{flalign}

The explicit spin wave function $\psi^{S(cc\bar c\bar c)}_{(S(cc)\otimes S(\bar c\bar c))}$ of $cc\bar c\bar c$ $[2](c_1c_2)\otimes[22](\bar c_3\bar c_4)$ configuration is listed as follows

\begin{flalign}\label{eq:spina6}
\psi^{S=0}_{(0\otimes0)}&=\frac 12(\uparrow \downarrow \bar \uparrow \bar \downarrow-\uparrow \downarrow \bar \downarrow \bar \uparrow-\downarrow \uparrow \bar \uparrow \bar \downarrow+\downarrow \uparrow \bar \downarrow \bar \uparrow)
\end{flalign}

The explicit spin wave functions $\psi^{S(cc\bar c\bar c)}_{(S(cc)\otimes S(\bar c\bar c))}$ of $cc\bar c\bar c$ $[11](c_1c_2)\otimes[211](\bar c_3\bar c_4)$ configuration are listed as follows

\begin{flalign}\label{eq:spina3}
\psi^{S=2}_{(1\otimes1)}&=\uparrow \uparrow \bar \uparrow \bar \uparrow, \nonumber \\
\psi^{S=1}_{(1\otimes1)}&=\frac 12(\uparrow \uparrow \bar \uparrow \bar \downarrow+\uparrow \uparrow \bar \downarrow \bar \uparrow-\uparrow \downarrow \bar \uparrow \bar \uparrow-\downarrow \uparrow \bar \uparrow \bar \uparrow), \nonumber \\
\psi^{S=0}_{(1\otimes1)}&=\frac 1{\sqrt 3}[\uparrow \uparrow \bar \downarrow \bar \downarrow-\frac 12(\uparrow \downarrow \bar \uparrow \bar \downarrow+\uparrow \downarrow \bar \downarrow \bar \uparrow+\downarrow \uparrow \bar \uparrow \bar \downarrow+\downarrow \uparrow \bar \downarrow \bar \uparrow) \nonumber \\
& +\downarrow \downarrow \bar \uparrow \bar \uparrow]
\end{flalign}

\indent The total spatial wave function of tetraquark, coupling among the $\sigma_1$, $\sigma_2$ and $\lambda$ harmonic oscillator wave functions, may take the general form,
\begin{eqnarray}\label{eqn::spatiala}
\psi_{NLM} &=&
\sum_{\{n_i,l_i\}}
 A(n_{\sigma_1},n_{\sigma_2},n_{\lambda},l_{\sigma_1},l_{\sigma_2},l_{\lambda}) \nonumber \\
&& \times \psi_{n_{\sigma_1}l_{\sigma_1}}(\vec\sigma_1\,) \otimes\psi_{n_{\sigma_2}l_{\sigma_2}}(\vec\sigma_2\,)\otimes\psi_{n_{\lambda}l_{\lambda}}(\vec\lambda\,)
\nonumber \\
&=& \sum_{\{n_i,l_i,m_i\}} C_{n_{\sigma_1}, l_{\sigma_1},m_{\sigma_1},n_{\sigma_2},l_{\sigma_2},m_{\sigma_2},n_{\lambda},l_{\lambda},m_{\lambda}} \nonumber \\
&& \times \psi_{n_{\sigma_1}l_{\sigma_1}m_{\sigma_1}}(\vec\sigma_1\,)\psi_{n_{\sigma_2}l_{\sigma_2}m_{\sigma_2}}(\vec\eta\,)\psi_{n_{\lambda}l_{\lambda}m_{\lambda}}(\vec\lambda\,) \nonumber \\
\end{eqnarray}
where $\psi_{n_{i}l_{i}m_{i}}$ are just harmonic oscillator wave functions and the sum $\{n_i,l_i\}$ is over $n_{\sigma_1},n_{\sigma_2},n_{\lambda}, l_{\sigma_1},l_{\sigma_2},l_{\lambda}$. $N$, $L$, and $M$ are respectively the total principle quantum number, total angular momentum, and magnetic quantum number of the tetraquark. One has $N= (2n_{\sigma_1}+ l_{\sigma_1})+(2n_{\sigma_2}+ l_{\sigma_2})+(2n_{\lambda}+l_{\lambda})$.

The complete bases of the tetraquarks are listed in Table \ref{three1} up to $N=10$, where $l_{\sigma_1}$, $l_{\sigma_2}$, and $l_{\lambda}$ are limited to $0$ only.

\begin{table*}[p]
\caption{\label{three1} The complete bases of tetraquark with quantum number, $N=2n$ and $L=M=0$.}
\begin{ruledtabular}
\begin{tabular}{@{}ll}
$\psi_{000}$ & $\psi_{0,0,0}(\vec\sigma_1\,)\psi_{0,0,0}(\vec\sigma_2\,)\psi_{0,0,0}(\vec\lambda\,)$ \\ [2pt]
$\psi_{200}$ & $\psi_{1,0,0}(\vec\sigma_1\,)\psi_{0,0,0}(\vec\sigma_2\,)\psi_{0,0,0}(\vec\lambda\,)$, $\psi_{0,0,0}(\vec\sigma_1\,)\psi_{1,0,0}(\vec\sigma_2\,)\psi_{0,0,0}(\vec\lambda\,)$, $\psi_{0,0,0}(\vec\sigma_1\,)\psi_{0,0,0}(\vec\sigma_2\,)\psi_{1,0,0}(\vec\lambda\,)$  \\ [2pt]
$\psi_{400}$ & $\psi_{2,0,0}(\vec\sigma_1\,)\psi_{0,0,0}(\vec\sigma_2\,)\psi_{0,0,0}(\vec\lambda\,)$, $\psi_{0,0,0}(\vec\sigma_1\,)\psi_{2,0,0}(\vec\sigma_2\,)\psi_{0,0,0}(\vec\lambda\,)$, $\psi_{0,0,0}(\vec\sigma_1\,)\psi_{0,0,0}(\vec\sigma_2\,)\psi_{2,0,0}(\vec\lambda\,)$, $\psi_{1,0,0}(\vec\sigma_1\,)\psi_{1,0,0}(\vec\sigma_2\,)\psi_{0,0,0}(\vec\lambda\,)$\\
\phantom{} & $\psi_{1,0,0}(\vec\sigma_1\,)\psi_{0,0,0}(\vec\sigma_2\,)\psi_{1,0,0}(\vec\lambda\,)$, $\psi_{0,0,0}(\vec\sigma_1\,)\psi_{1,0,0}(\vec\sigma_2\,)\psi_{1,0,0}(\vec\lambda\,)$ \\[2pt]
$\psi_{600}$ & $\psi_{3,0,0}(\vec\sigma_1\,)\psi_{0,0,0}(\vec\sigma_2\,)\psi_{0,0,0}(\vec\lambda\,)$, $\psi_{2,0,0}(\vec\sigma_1\,)\psi_{1,0,0}(\vec\sigma_2\,)\psi_{0,0,0}(\vec\lambda\,)$,$\psi_{2,0,0}(\vec\sigma_1\,)\psi_{0,0,0}(\vec\sigma_2\,)\psi_{1,0,0}(\vec\lambda\,)$, $\psi_{1,0,0}(\vec\sigma_1\,)\psi_{2,0,0}(\vec\sigma_2\,)\psi_{0,0,0}(\vec\lambda\,)$, \\ [2pt]
\phantom{} & $\psi_{1,0,0}(\vec\sigma_1\,)\psi_{0,0,0}(\vec\sigma_2\,)\psi_{2,0,0}(\vec\lambda\,)$, $\psi_{0,0,0}(\vec\sigma_1\,)\psi_{3,0,0}(\vec\sigma_2\,)\psi_{0,0,0}(\vec\lambda\,)$, $\psi_{0,0,0}(\vec\sigma_1\,)\psi_{2,0,0}(\vec\sigma_2\,)\psi_{1,0,0}(\vec\lambda\,)$, $\psi_{0,0,0}(\vec\sigma_1\,)\psi_{1,0,0}(\vec\sigma_2\,)\psi_{2,0,0}(\vec\lambda\,)$, \\
\phantom{} & $\psi_{0,0,0}(\vec\sigma_1\,)\psi_{0,0,0}(\vec\sigma_2\,)\psi_{3,0,0}(\vec\lambda\,)$, $\psi_{1,0,0}(\vec\sigma_1\,)\psi_{1,0,0}(\vec\sigma_2\,)\psi_{1,0,0}(\vec\lambda\,)$\\
$\psi_{800}$ & $\psi_{4,0,0}(\vec\sigma_1\,)\psi_{0,0,0}(\vec\sigma_2\,)\psi_{0,0,0}(\vec\lambda\,)$, $\psi_{3,0,0}(\vec\sigma_1\,)\psi_{1,0,0}(\vec\sigma_2\,)\psi_{0,0,0}(\vec\lambda\,)$, $\psi_{3,0,0}(\vec\sigma_1\,)\psi_{0,0,0}(\vec\sigma_2\,)\psi_{1,0,0}(\vec\lambda\,)$, $\psi_{2,0,0}(\vec\sigma_1\,)\psi_{2,0,0}(\vec\sigma_2\,)\psi_{0,0,0}(\vec\lambda\,)$,  \\ [2pt]
\phantom{} & $\psi_{2,0,0}(\vec\sigma_1\,)\psi_{1,0,0}(\vec\sigma_2\,)\psi_{1,0,0}(\vec\lambda\,)$, $\psi_{2,0,0}(\vec\sigma_1\,)\psi_{0,0,0}(\vec\sigma_2\,)\psi_{1,0,0}(\vec\lambda\,)$, $\psi_{1,0,0}(\vec\sigma_1\,)\psi_{3,0,0}(\vec\sigma_2\,)\psi_{0,0,0}(\vec\lambda\,)$, $\psi_{1,0,0}(\vec\sigma_1\,)\psi_{2,0,0}(\vec\sigma_2\,)\psi_{1,0,0}(\vec\lambda\,)$\\
\phantom{} & $\psi_{1,0,0}(\vec\sigma_1\,)\psi_{1,0,0}(\vec\sigma_2\,)\psi_{2,0,0}(\vec\lambda\,)$, $\psi_{1,0,0}(\vec\sigma_1\,)\psi_{0,0,0}(\vec\sigma_2\,)\psi_{3,0,0}(\vec\lambda\,)$, $\psi_{0,0,0}(\vec\sigma_1\,)\psi_{4,0,0}(\vec\sigma_2\,)\psi_{0,0,0}(\vec\lambda\,)$, $\psi_{0,0,0}(\vec\sigma_1\,)\psi_{3,0,0}(\vec\sigma_2\,)\psi_{1,0,0}(\vec\lambda\,)$, \\
\phantom{} & $\psi_{0,0,0}(\vec\sigma_1\,)\psi_{2,0,0}(\vec\sigma_2\,)\psi_{2,0,0}(\vec\lambda\,)$, $\psi_{0,0,0}(\vec\sigma_1\,)\psi_{1,0,0}(\vec\sigma_2\,)\psi_{3,0,0}(\vec\lambda\,)$, $\psi_{0,0,0}(\vec\sigma_1\,)\psi_{0,0,0}(\vec\sigma_2\,)\psi_{4,0,0}(\vec\lambda\,)$\\
$\psi_{1000}$ & $\psi_{5,0,0}(\vec\sigma_1\,)\psi_{0,0,0}(\vec\sigma_2\,)\psi_{0,0,0}(\vec\lambda\,)$, $\psi_{4,0,0}(\vec\sigma_1\,)\psi_{1,0,0}(\vec\sigma_2\,)\psi_{0,0,0}(\vec\lambda\,)$, $\psi_{4,0,0}(\vec\sigma_1\,)\psi_{0,0,0}(\vec\sigma_2\,)\psi_{1,0,0}(\vec\lambda\,)$, $\psi_{3,0,0}(\vec\sigma_1\,)\psi_{2,0,0}(\vec\sigma_2\,)\psi_{0,0,0}(\vec\lambda\,)$,\\
\phantom{} & $\psi_{3,0,0}(\vec\sigma_1\,)\psi_{1,0,0}(\vec\sigma_2\,)\psi_{1,0,0}(\vec\lambda\,)$, $\psi_{3,0,0}(\vec\sigma_1\,)\psi_{0,0,0}(\vec\sigma_2\,)\psi_{2,0,0}(\vec\lambda\,)$, $\psi_{2,0,0}(\vec\sigma_1\,)\psi_{3,0,0}(\vec\sigma_2\,)\psi_{0,0,0}(\vec\lambda\,)$, $\psi_{2,0,0}(\vec\sigma_1\,)\psi_{2,0,0}(\vec\sigma_2\,)\psi_{1,0,0}(\vec\lambda\,)$,\\
\phantom{} & $\psi_{2,0,0}(\vec\sigma_1\,)\psi_{1,0,0}(\vec\sigma_2\,)\psi_{1,0,0}(\vec\lambda\,)$, $\psi_{2,0,0}(\vec\sigma_1\,)\psi_{0,0,0}(\vec\sigma_2\,)\psi_{3,0,0}(\vec\lambda\,)$, $\psi_{1,0,0}(\vec\sigma_1\,)\psi_{4,0,0}(\vec\sigma_2\,)\psi_{0,0,0}(\vec\lambda\,)$, $\psi_{1,0,0}(\vec\sigma_1\,)\psi_{3,0,0}(\vec\sigma_2\,)\psi_{1,0,0}(\vec\lambda\,)$,\\
\phantom{} & $\psi_{1,0,0}(\vec\sigma_1\,)\psi_{2,0,0}(\vec\sigma_2\,)\psi_{2,0,0}(\vec\lambda\,)$, $\psi_{1,0,0}(\vec\sigma_1\,)\psi_{1,0,0}(\vec\sigma_2\,)\psi_{3,0,0}(\vec\lambda\,)$, $\psi_{1,0,0}(\vec\sigma_1\,)\psi_{0,0,0}(\vec\sigma_2\,)\psi_{4,0,0}(\vec\lambda\,)$, $\psi_{0,0,0}(\vec\sigma_1\,)\psi_{5,0,0}(\vec\sigma_2\,)\psi_{0,0,0}(\vec\lambda\,)$,\\
\phantom{} & $\psi_{0,0,0}(\vec\sigma_1\,)\psi_{4,0,0}(\vec\sigma_2\,)\psi_{1,0,0}(\vec\lambda\,)$, $\psi_{0,0,0}(\vec\sigma_1\,)\psi_{3,0,0}(\vec\sigma_2\,)\psi_{2,0,0}(\vec\lambda\,)$, $\psi_{0,0,0}(\vec\sigma_1\,)\psi_{2,0,0}(\vec\sigma_2\,)\psi_{3,0,0}(\vec\lambda\,)$, $\psi_{0,0,0}(\vec\sigma_1\,)\psi_{1,0,0}(\vec\sigma_2\,)\psi_{4,0,0}(\vec\lambda\,)$,\\
\phantom{} & $\psi_{0,0,0}(\vec\sigma_1\,)\psi_{0,0,0}(\vec\sigma_2\,)\psi_{5,0,0}(\vec\lambda\,)$
 \\ [2pt]
\end{tabular}
\end{ruledtabular}
\end{table*}

\bibliography{arXiv2020}

\end{document}